\documentclass[a4paper,11pt]{article}
\usepackage{fullpage}
\usepackage{xcolor}
\usepackage{graphicx}
\usepackage{amssymb}
\usepackage{amsthm}
\usepackage{amsmath}
\usepackage{stmaryrd} 

\begin{document}

\newcommand*\luk[2]{{\color{red}{#2}}}%
\newcommand*\win[2]{{\color{blue}{#2}}}%
\newcommand*\Paragraph{\smallskip}%
\newcommand*\Startsec{\vspace*{0.75\jot}\noindent}%
\newcommand*\vph{$\vphantom{\int\nolimits^X}$}%

\title{Fast stray field computation on tensor grids}

\author{L.~Exl\thanks{\texttt{Corresponding author, lukas.exl@fhstp.ac.at}}, S.~Bance, M.~Gusenbauer, F.~Reichel, T.~Schrefl \\[0.1cm] University of Applied Sciences, \\Department of Technology, 
Matthias Corvinus-Stra\ss e 15, A-3100 St.\,P{\"o}lten, Austria\\[0.2cm] 
W.~Auzinger \\ Vienna University of Technology, \\[0.1cm] Institute for Analysis and Scientific Computing , A-1040 Vienna, Austria}

\maketitle

\begin{abstract}
A direct integration algorithm is described to compute the magnetostatic field and energy
for given magnetization distributions on not necessarily uniform tensor grids.
We use an analytically-based tensor approximation approach for function-related tensors,
which reduces calculations to multilinear algebra operations. The algorithm scales with $N^{4/3}$ for $N$ computational cells used and with $N^{2/3}$ (\textit{sublinear}) when magnetization is given in canonical
tensor format. In the final section we confirm our theoretical results
concerning computing times and accuracy by means of numerical examples.
\end{abstract}

\Paragraph
Micromagnetics \quad Stray field \quad Tensor grids \quad Low-rank \quad Canonical format \quad Tucker tensor

\section{Introduction}
\label{intro}

\Startsec
Computation of the magnetostatic field is the most time-consuming aspect in micromagnetic simulations. This is usually done by
evaluating the magnetostatic scalar potential, which
involves the solution of a Poisson equation, e.g., by means of
a hybrid FEM/BEM method, see e.g.~\cite{kronmuller_handbook_2007}.
Alternatively, direct computation by discretization of a volume integral formulation of the scalar potential normally scales
with the total number of computational cells squared, i.e.~$O(n^6)$ for $n^3$ cells. Several techniques
have been introduced in the literature
to reduce computational costs, e.g.\
the fast multipole method (combined with FFT), \cite{blue_using_1991}, \cite{FFTM}, NG methods, \cite{NG_2009}, scaling linearly, i.e. $O(n^{3})$ and H-matrix techniques, \cite{pop_2004}, with almost linear complexity, i.e.~$O(n^3\log n)$.  
Recent developments show sublinear compression properties, i.e.~$O\big(n^2 \big)$,
both for storage requirements and computational complexity by applying
multilinear algebra approximation techniques to the
demagnetizing and magnetization tensors \cite{goncharov_kronecker_2010}.
For this purpose the magnetization tensor has to be represented or approximated in
canonical tensor format~\cite{kolda_tensor_2009}, and a non-regular grid cannot be used.
Some further difficulties appear when a purely algebraic approach is used for tensor approximation,
e.g., the best low-rank approximation problem is not well-posed and globally convergent algorithms
do not exist so far~\cite{de_silva_tensor_2008}, \cite{kolda_tensor_2009}.

\Paragraph
In Sec.~\ref{method}, we present an analytically-based tensor approximation method,
which can be generally used for a special class of function-related tensors~\cite{hackbusch_low-rank_2005}.
Calculation of the scalar potential and the stray field reduces to multilinear algebra operations,
which can be implemented efficiently using
optimized libraries~\cite{oseledets_linear_2009}, \cite{bader_algorithm_2006}, \cite{bader_efficient_2008}. 
Magnetization can also be treated as low-rank tensors and the data-sparse format is preserved by the method, which scales almost linearly, i.e.~$O(n^4)$ in the general case and \textit{sublinear},
i.e., $O(n^2)$ for specially structured magnetization tensors, e.g.\ in CP format (see~\ref{tensors3})
which was also used in~\cite{goncharov_kronecker_2010}.
Using canonical tensor formats could open up new possibilities for solving the variational model by Landau-Lifschitz for stationary micromagnetic phenomena efficiently by projection methods (e.g., CG or GMRES) for linear systems in tensor format,
recently introduced in \cite{tensorLGS}.\\
Because of the sublinear scaling of tensor-grid methods, micromagnetic methods that compute magnetization dynamics or hysterisis properties (whereby the magnetization is in the CP format) have a high potential for solving large scale 
engineering problems. The aim of this work is to provide a key-building block of such an algorithm: The computation of the magnetostatic energy from magnetization distriubutions given in CP format.

\Paragraph
In Sec.~\ref{Num} we test our algorithm by computing the magnetostatic potential, field and energy of
hexahedral ferromagnetic bodies for different given magnetizations.

\section{Method}
\label{method}

\subsection{Analytical preparations}
\label{analy}

\Startsec
The magnetostatic scalar potential in a ferromagnetic body $\Omega \subset \mathbb{R}^3$ induced by a
given magnetization distribution $\textbf{M}$ is usually given by the formula \cite{jackson_classical_1998}
\begin{align} \label{scpot2}
\phi(\vec{x}) = -\frac{1}{4\,\pi} \int_{\Omega} \frac{\nabla \cdot \textbf{M}(\vec{x}^{\,\prime})}
                                                   {\left|\vec{x} - \vec{x}^{\,\prime} \right|}\,d^{\,3} x^{\prime}
                + \frac{1}{4\,\pi} \oint_{\partial \Omega} \frac{\vec{n}^{\prime} \cdot \textbf{M}(\vec{x}^{\,\prime})}
                                                {\left|\vec{x} - \vec{x}^{\,\prime} \right|}\,da^{\prime},
\end{align}
and the stray field then reads
\begin{equation} \label{field}
        \textbf{H}_d = - \nabla\phi.
\end{equation}
We aim for computing the scalar potential by means of multilinear tensor operations and therefore
prefer discretizing a volume integral instead of the formulation from Eq.~\eqref{scpot2}.
Integration by parts leads to
\begin{align} \label{scpot}
 		\phi(\vec{x}) = \frac{1}{4\,\pi}  \int_{\Omega} \textbf{M}(\vec{x}^{\,\prime}) \cdot
                        \frac{\vec{x} - \vec{x}^{\,\prime}}
                              {\left|\vec{x} - \vec{x}^{\,\prime} \right|^3}\,d^{\,3} x^{\prime}.
\end{align}
We denote the three volume integrals in Eq.~\eqref{scpot} by
\begin{align} \label{intcomp}
		I^{(p)}(\vec{x}) = \int_{\Omega} M^{(p)}(\vec{x}^{\,\prime})\,                           \frac{x^{(p)} - {x^{\prime}}^{\,(p)}}
                                {\left|\vec{x} - \vec{x}^{\,\prime} \right|^3}\,d^{\,3} x^{\prime},
\end{align}
for each of the components $M^{(p)}$, $p=1 \hdots 3$,
of the magnetization $\textbf{M}$, so Eq.~\eqref{scpot} reads
\begin{align} \label{scpot3}
		\phi(\vec{x}) = \frac{1}{4\,\pi} \sum_{p=1}^{3} I^{(p)}(\vec{x}).
\end{align}
As a first step, in order to get rid of the singularities at
$\vec{x}^{\,\prime} = \vec{x}$,
we represent the integral kernel in \eqref{intcomp} as an integral of a \textit{Gaussian function} by the formula
\begin{align} \label{transform}
\frac{1}{\rho^{\frac{3}{2}}} = \frac{2}{\sqrt{\pi}} \int_{\mathbb{R}} \tau^2\,e^{-\tau^2\rho}\,d\tau.
\end{align}
So one has, for $\rho=|\vec{x}-\vec{x}^{\,\prime}|^2$,
\begin{align} \label{gauss}
I^{(p)}(\vec{x}) = \frac{2}{\sqrt{\pi}} \int_{\mathbb{R}} \tau^2 \int_{\Omega}
                   e^{-\tau^2\,|\vec{x}-\vec{x}^{\,\prime}|^2}M^{(p)}(\vec{x}^{\,\prime})(x^{(p)} - {x^{\prime}}^{\,(p)})
                  \,d^{\,3} x^{\prime}\,d\tau.
\end{align}
Eq.\eqref{gauss} reduces the computation to independent integrals along each principal direction. As we will see later, this results in a reduction of computational effort from $O(N^2)$ to $O(N^{4/3})$.

\Paragraph
The rest of the paper is organized as follows. First we discuss the $\tau$\,-\,integration in
Eq.~\eqref{gauss} via \textit{sinc-quadrature}.
Then we consider the spatial discretization of the resulting quadrature approximation on tensor grids and
discuss its computational realization.
In Sec.~\ref{Num} we present some numerical results on computational complexity and accuracy for different given magnetizations.

\subsection{$\tau$\,-\,Integration}
\label{tauint}

\Startsec
Although the singularity is gone, the numerical treatment of \eqref{gauss} is not
straightforward.
Small values of $\rho$ have a dispersive effect on the integrand in \eqref{transform},
so one has to distribute the quadrature nodes over a wide range for accurate approximation
of the kernel function $1/\rho^{3/2}$.
These values of $\rho$ correspond to a small grid size, which is again essential
for an accurate approximation of the magnetic scalar potential.
Therefore one has to use a quadrature rule that is robust with respect to small values of $\rho$.

\Paragraph
For Gaussian quadrature, weights and nodes can be computed by
solving an eigenvalue problem of symmetric tridiagonal type.
For a larger number of quadrature points one uses the QR algorithm,
which scales linearly in the number of quadrature terms.
On the other hand, using Gaussian quadrature formulas on a finite subinterval $[0,A]$,
or Gauss-Laguerre quadrature over the infinite interval in~\eqref{transform}
fail in terms of achieving a sufficiently good representation of the kernel function for small values of $\rho$.

\Paragraph
In \cite{juselius_parallel_2007} an integral representation for the \textit{Newton potential},
i.e., $1/\rho$, was used to compute the electrostatic potential for
given Gaussian distributed charges in whole space.
The $\tau$\,-\,integration was performed using the
Gauss-Legendre formula on logarithmically scaled blocks of the interval $[0,10^4]$ using a total of $120$ quadrature points.
Due to geometrical refinement against zero, the functional $1/\rho$ is well described in this region.

\Paragraph
Here we use the exponentially convergent \textit{sinc-quadrature}
for numerical integration of the integral~\,\cite{hackbusch_low-rank_2005}.
This method shows better approximation properties than
the Gauss-Legendre formula for much fewer quadrature terms.

\Paragraph
Abbreviating notation and exploiting symmetry in $\tau$, the $I^{(p)}$ take the form
\begin{align}\label{Ip}
 I^{(p)}(\vec{x}) = \frac{2}{\sqrt{\pi}}\int_{0}^{\infty}\tau^2\,2\,F^{(p)}(\vec{x},\tau)\,d\tau,
\end{align}
where $F^{(p)}$ stands for the $\Omega$\,-\,integral in~\eqref{gauss}. In order to guarantee the above mentioned exponential convergence rate 
we perform an integral transform on Eq.~\eqref{Ip}, i.e. $\tau = \sinh(t)$, and apply numerical integration afterwards, which gives
\begin{align} \label{quad}
		 \int_{0}^{\infty}\tau^2\,2\,F^{(p)}(\vec{x},\tau)\,d\tau  
\approx \sum_{l=1}^{R}\omega_{l}\,\sinh(t_{l})^2\,G^{(p)}(\vec{x},t_{l}),
\end{align}
where $(t_l,\omega_l)$ are the nodes and weights of the
underlying quadrature, and $G^{(p)}(\vec{x},t) = 2\,F^{(p)}(\vec{x},\sinh(t))$.

\Paragraph
To apply \textit{sinc-quadrature}\, we use 
the $R+1$ nodes and weights 
given by
\begin{align}
   t_l = l h_R
\end{align}
and
\begin{align}
\omega_l = \left\{\begin{array}{l l}
                   h_R                 & l=0, \\*[\jot]
                   2\,h_{R}\cosh(t_{l}) & l>0,
	\end{array}\right.
\end{align}
with $h_R = c_0\,\ln(R)/R$ for some appropriate
$c_0$, see Proposition $2.1$ in \cite{hackbusch_low-rank_2005}.	

Tab.~\ref{table_sinc} shows the average absolute and relative errors due to sinc-approximation of the functional~\eqref{transform}
for $10^5$ equidistantly chosen $\rho$-values of the interval $[5e-05,1e-02]$ (which corresponds to a mesh size of $10^{-1}$ up to $200^{-1}$ in a uniform tensor grid, see Sec.~\ref{discret}). 
The left three columns show accuracy for different values of the parameter $c_0$ and $R=50$, right columns show dependence of the number of quadrature terms and $c_0 = 1.85$.

\begin{table}
\caption{Average abs./rel. errors of sinc-quadrature on $\rho$-intervall $[5e-05,1e-02]$. Left: $c_0$-dependence of approximation of \eqref{transform} $(R=50)$. Right: $R$-dependence of approximation of \eqref{transform} $(c_0 = 1.85)$.}\label{table_sinc} 

\begin{center}
\begin{tabular}{|c|c|c||c|c|c|}\hline
$c_0 $  &  abs. error  & rel. error & $R$ & abs. error   & rel. error\\ \hline\hline
$1.70$ &  8.3\,$e-$01 & 3.4\,$e-$07 & $35$ &  1.6\,$e+$00  & 6.5\,$e-$07\vph \\ \hline
$1.75$ &  8.4\,$e-$03 & 3.3\,$e-$09 & $40$ &  7.4\,$e-$03  & 2.9\,$e-$09\vph \\ \hline
$1.80$ &  8.5\,$e-$06 & 3.3\,$e-$12 & $45$ &  4.7\,$e-$06  & 3.2\,$e-$12\vph \\ \hline
$1.85$ &  3.4\,$e-$09 & 1.2\,$e-$13 & $50$ &  3.4\,$e-$09  & 1.2\,$e-$13\vph \\ \hline
$1.90$ &  7.8\,$e-$09 & 3.0\,$e-$13 & $55$ &  2.8\,$e-$10  & 1.1\,$e-$14\vph \\ \hline
$1.95$ &  1.7\,$e-$08 & 6.7\,$e-$13 & $60$ &  2.5\,$e-$11  & 9.5\,$e-$16\vph \\ \hline
$2.00$ &  3.9\,$e-$08 & 1.5\,$e-$12 & $65$ &  3.3\,$e-$12  & 1.2\,$e-$16\vph \\ \hline
$2.05$ &  8.2\,$e-$08 & 3.0\,$e-$12 & $70$ &  4.4\,$e-$12  & 8.9\,$e-$17\vph \\ \hline
$2.10$ &  1.7\,$e-$07 & 6.3\,$e-$12 & $75$ &  2.3\,$e-$12  & 9.3\,$e-$17\vph \\ \hline
$2.15$ &  3.2\,$e-$07 & 1.2\,$e-$11 & $80$ &  2.4\,$e-$12  & 9.1\,$e-$17\vph \\ \hline
\end{tabular}
\end{center}
\end{table}

For our numerical experiments in Sec.~\ref{Num}, we use $c_0 = 1.85$,
which gives a sufficiently good description of the functional~\eqref{transform}.\\
One could think about optimizing $c_0$ by minimizing the number of quadrature terms for a given accuracy of the description of the functional \eqref{transform}, 
but we do not address possible algorithmic realizations of this topic in this paper.


\subsection{Discretization on tensor grids}
\label{discret}

\Startsec
For the sake of simplicity let us first assume a uniformly discretized cube $\Omega$,
where the tensor product grid consists of $n^3$ subcubes $\Omega_{\textbf{j}}, \ \textbf{j}:=(j_1 , j_2 , j_3)$.
We make the assumption of constant magnetization for each spatial component $p = 1 \hdots 3$ in each subcube, i.e.
\begin{align} \label{magnet}
\textbf{M}^{(p)} = \sum_{\textbf{j}} m_{\textbf{j}}^{(p)} \chi_{\Omega_{\textbf{j}}},
\end{align}
where $\chi_{\Omega_{\textbf{j}}} = \chi_{\Omega_{j_1}} \chi_{\Omega_{j_2}} \chi_{\Omega_{j_3}}$
is the $3$-d characteristic function of the subcube $\Omega_{\textbf{j}}$,
and $m_{\textbf{j}}^{(p)}$ the components of the $3$-d
\textit{$p$-component magnetization tensor} $\textbf{M}^{(p)}$.

\Paragraph
The computational realization  of the quadrature approximation~\eqref{quad} to Eq.~\eqref{gauss} requires
evaluation of the scalar potential at the center point
$x_{\textbf{i}}^{c} = (x_{i_1}^{c}, x_{i_2}^{c}, x_{i_3}^{c})$ of each field cell.
To this end we substitute Eq.~\eqref{magnet} into the function $G^{(p)}$ of Eq.~\eqref{quad}.
This leads to
\begin{align} \label{tuck1}
     G^{(p)}(x_{\textbf{i}}^{c},t_{l}) =
     \sum_{\textbf{j}} m_{\textbf{j}}^{(p)}
     \prod_{q=1}^{3}
     \int_{\Omega} g^{(q)}(x_{i_{q}}^{c}, x^{\prime}, t_{l})\,\chi_{\Omega_{j_{q}}}(x^{\prime})\,dx^{\prime},
\end{align}
where
\begin{align}
  g^{(q)}(\alpha,\alpha^{\prime}, \tau) :=
      \left\{\begin{array}{l l}
	     \exp(-\sinh(\tau)^2\,(\alpha - \alpha^{\prime})^2)                              & q \neq p, \\*[\jot]
	     (\alpha - \alpha^{\prime}) \exp(-\sinh(\tau)^2\,(\alpha - \alpha^{\prime} )^2)  & q = p.
	  \end{array}\right.
\end{align}
The three integrals in Eq.~\eqref{tuck1} define $(n \times n)$\,-\,matrices, i.e.
%
\begin{align} \label{Dmat}
  d_{i_q,j_q}^{\,l} := & \int_{\Omega_{j_{q}}} g(x_{i_{q}}^{c}, x^{\prime}, t_{l})\,dx^{\prime}, \\
  D_{q}^{\,l} := & \big( d_{i_q\,j_q}^{\,l} \big).
\end{align}
%
So we have a Tucker representation
of the function $G^{(p)}$ (see~\ref{tensors3}), i.e.,
\begin{align} \label{tuck2}
G^{(p)}(x_{\textbf{i}}^{c},t_{l})
& = \sum_{\textbf{j}} m_{j_1\,j_2\,j_3}^{(p)} \ d_{i_1\,j_1}^{\,l} \ d_{i_2\,j_2}^{\,l} \ d_{i_3\,j_3}^{\,l} \\
& = \textbf{M}^{(p)} \times_1 D_{1}^{\,l} \times_2 D_{2}^{\,l} \times_3 D_{3}^{\,l},
\end{align}
with the core tensor $\textbf{M}^{(p)}$.

\subsection{Magnetization in CP format}
\label{CPmag}

\Startsec
The main goal of this paper is the development of an algorithm for magnetostatics that allows the treatment of magnetization in tensor low-rank formats and also preserves this format. 
In the following, we can see how the preparations of the previous sections make it possible to compute the potential for CP-magnetization.

Combining Eqs. \eqref{scpot}, \eqref{quad} and \eqref{tuck2} yields the scalar potential at the center points
\begin{align}\label{sumpot}
  \mathbb{R}^{n \times n \times n} \ni \Phi = \frac{1}{2 \pi^{3/2}} \sum_{p=1}^{3} \sum_{l=1}^{R} \omega_{l}\,\sinh(t_{l})^2\,\textbf{M}^{(p)} \times_1 D_{1}^{\,l} \times_2 D_{2}^{\,l} \times_3 D_{3}^{\,l}. 
\end{align}
Assuming $\textbf{M}^{(p)} \in \mathcal{C}_{n,r_p}$, see \ref{tensors3}, shows \eqref{sumpot} to be in canonical format as well, i.e.
\begin{align}\label{cppot}
 \Phi \propto \sum_{p=1}^{3} \sum_{l=1}^{R} \llbracket \lambda^{(p)};\,D_{1}^{\,l} M_{1}^{(p)},D_{2}^{\,l} M_{2}^{(p)},D_{3}^{\,l} M_{3}^{(p)} \rrbracket \in \mathcal{C}_{n,R\sum_{p=1}^{3} r_p}, 
\end{align}
where $M_{q}^{(p)} \in \mathbb{R}^{n \times r_{p}}, \, \lambda^{(p)} \in \mathbb{R}^{r_{p}},\, q,p = 1 \hdots 3$ and the factors $\omega_l \, \sinh(t_{l})^2$ are absorbed by the weight vectors $\lambda^{(p)}$.\\

\subsection{Non-uniform grids}
\label{nonuni}

\Startsec
Let now $\Omega$ be a non-uniformly partitioned cube, i.e.
\begin{align}
  \Omega = \bigcup_{\textbf{j}\,\in\,\textbf{J}}\,\Omega_{\textbf{j}}, & \qquad
  \Omega_{\textbf{j}} = \prod_{i=1}^{3}\,\Omega_{j_{i}},
\end{align}
where $\textbf{J} \subset \mathbb{N}_{+}^{3}$ is a set of multiindices.

\Paragraph
For the following we define $n_{p} := \max_{\,\textbf{j}\,\in\,\textbf{J}} j_{p}$, and the extension $\Tilde{\textbf{J}}$ of $\textbf{J}$ by
\begin{align}
 \Tilde{\textbf{J}} := \left\{\,\textbf{j} \mid j_{p} = 1 \hdots n_{p}, ~p=1 \hdots 3\,\right\}.
\end{align}
Similarly, the magnetization \eqref{magnet} is then given by
\begin{align}
 \textbf{M}^{(p)} = \sum_{\textbf{j}\,\in\,\Tilde{\textbf{J}} } m_{\textbf{j}}^{(p)} \chi_{\Omega_{\textbf{j}}},
\end{align}
where we set $m_{\textbf{j}}^{(p)} = 0$ for $\textbf{j} \in \Tilde{\textbf{J}} \setminus \textbf{J}$.

\Paragraph
The \textit{$p$-component magnetization tensor} is thus formally an element of
$\mathbb{R}^{n_1 \times n_2 \times n_3}$, and the remaining approach is similar to that in Sec.~\ref{discret}.
The matrices in \eqref{Dmat} are now of different dimensions
for each component $q$, i.e.,  $D_{q}^{\,l} \in \mathbb{R}^{n_q \times n_q}$.
The magnetization tensors might now be sparse, so for the mode-multiplication in~\eqref{tuck2}
this can be taken into account to reduce computational costs \cite{bader_efficient_2008}.

\subsection{Computational issues}
\label{comp}

\Startsec
For evaluating \eqref{tuck2} we use \textit{Gauss-Legendre} quadrature with not more than $50$ terms
for the Gaussian integrals in \eqref{Dmat}.

\Paragraph
In the general case one \textit{$q$-mode matrix multiplication} for the $p$-component tensor $\textbf{M}^{(p)}$ in \eqref{tuck2} requires
$n$ matrix-matrix products and thus can be performed in $O(n^4)$ operations \cite{bader_efficient_2008}.
Since for mode multiplication the $n$ matrix products along the corresponding mode are independent,
it is easy to perform it in parallel, which reduces computation time, when using current state of the art computer architecture.

\Paragraph
When further special structure for the $p$-component magnetization tensors or the matrices in \eqref{tuck2}
is given, costs reduce significantly. If $\textbf{M}^{(p)}$ is assumed to be in canonical format, i.e. $\textbf{M}^{(p)} \in \mathcal{C}_{n,r}$,
(as in \cite{goncharov_kronecker_2010}), 
costs for the \textit{$q$-mode matrix multiplication} reduce to $O(r\,n^2)$ (see Eq.~\eqref{cpviatuck} or \cite{bader_efficient_2008}),
where $r$ stands for the \textit{Kronecker rank}. \\
Thus, Eq.~\eqref{tuck2} scales $O(n^4)$ in the general case respectively $O(n^2)$ for canonical magnetization and is computed for each $l$,
which does not depend on the grid spacing for robust $\tau$\,-\,quadrature described above.
Therefore the number of $\tau$\,-\,quadrature terms have no influence on computational costs with respect to $n$.
Summarizing, the total operation count for \eqref{sumpot} and \eqref{cppot} becomes approximately
\begin{align}\label{operation_count}
 3R(g/n^2 + 3)n^4 & \quad \text{for \eqref{sumpot}, \, and}
\end{align}
\begin{align}\label{operation_count2}
 3R(g+\sum_{p=1}^{3}r_{p})n^2  & \quad \text{for \eqref{cppot}},
\end{align}
where $g$ and $R$ denote the number of Gaussian- and sinc-quadrature terms, respectively.

\Paragraph
Although we do not intend to create an alternative to existing algorithms for the computation of pairwise charge or mass interactions with linear complexity in $N=n^3$ (e.g. FMM \cite{FMM_1997} or NG methods \cite{NG_2009}) rather than presenting a way for the computation of magnetostatic potentials/fields for data-sparse CP-magnetization, we briefly want to compare in terms of operation counts.\\
Asymptotic operation counts for different versions of FMM can be found in the literature. For instance, in \cite{FMM_1997} the fastest version in three dimensions  (exponential translation) counts approximately $200N p+ 3.5N p^2$ operations, where $N$ is the total number of cells and $p = \log_{\sqrt{3}}{1/\epsilon}$ (number of multipoles), for given accuracy $\epsilon$ of the multipole expansion (the average amount of particles in one box at the finest level is here $s = 2p$ for sake of convenience). 
Chosing $p=30$, for ten digits of accuracy, gives approximately $10^4 N$ operation for the calculation of the magnetic scalar potential due to pairwise interactions. 
As can be easily recalculated, this clearly outperforms \eqref{operation_count} (e.g. for grids larger than $N = 20^3$, with $R = g = 50$), but is itself outperformed by \eqref{operation_count2} for any grid size, when assuming $r_{p} \leq n$.\\
At this point it is also worth mentioning that in the case of CP-magnetization with $r_{p} \leq n$ the storage for the magnetization tensors is compressed by a factor of $\sum_{p=1}^{3}r_{p}/n^2 \leq 3/n$. We also want to refer to Sec.\ref{Num}, where the Eqs. \eqref{operation_count} and \eqref{operation_count2} are confirmed by numerical examples, as well as, the accuracy of the introduced approach is discussed.

\Paragraph
The resulting $I^{(p)}$ are $3$-d tensors defined on the tensor product grid.
Once the potential has been computed
, one has to perform discrete differentiation
to obtain the field~\eqref{field}. This can be done by \textit{$q$-mode sparse matrix multiplication},
which scales $O(n^3)$ for \eqref{sumpot}, as does \textit{$q$-mode vector multiplication}, and $O(n^2)$ for the canonical version \eqref{cppot}, see \cite{bader_efficient_2008}.
Here we use a sparse finite-difference matrix corresponding to a three-point
finite-difference approximation of order~$2$ for the first derivative.
Assuming a (not necessarily uniform) mesh in one spatial direction $p$, e.g., with mesh sizes $h_j, j = 1 \hdots n$,
one has to use general finite-difference approximations.
Since the potential is only given at the center points,
we first denote by $\tilde{h}_j := (h_j + h_{j+1})/2,~j=1\hdots n-1$,
the distance between successive midpoints.
For interior points the corresponding
second order centered finite-difference approximations are given by
\begin{align}
 &\alpha_{0}^{k,l} f(x - \tilde{h}_{k}) + \alpha_{1}^{k,l} f(x) + \alpha_{2}^{k,l} f(x + \tilde{h}_{l}) = f^{\prime}(x) + O(\tilde{h}_{k}\,\tilde{h}_{l}), ~~\text{with} \nonumber \\
 &\alpha_{0}^{k,l} = -\frac{\tilde{h}_{l}}{\tilde{h}_{k}(\tilde{h}_{k} + \tilde{h}_{l})}, \quad
  \alpha_{1}^{k,l} = \frac{\tilde{h}_{l} - \tilde{h}_{k}}{\tilde{h}_{k}\,\tilde{h}_{l}}, \quad
  \alpha_{2}^{k,l} = \frac{\tilde{h}_{k}}{\tilde{h}_{l}(\tilde{h}_{k} + \tilde{h}_{l})},
\end{align}
where $ \tilde{h}_{k},\tilde{h}_{l} $ are the distances to the left and right neighbor of a midpoint $ x $, respectively.
For the boundaries we use the analogous one-sided scheme
\begin{align}
 &\beta_{0}^{k,l} f(x) + \beta_1^{k,l} f(x + \tilde{h}_{k}) + \beta_2^{k,l} f(x + \tilde{h}_{k} + \tilde{h}_{l}) =  f^{\prime}(x) + O(\tilde{h}_{k}(\tilde{h}_{k} + \tilde{h}_{l})), \quad \text{with} \nonumber \\
 &\beta_{0}^{k,l} = -\frac{2\,\tilde{h}_{k}+\tilde{h}_{l}}{\tilde{h}_{k}(\tilde{h}_{k} + \tilde{h}_{l})}, \quad
  \beta_1^{k,l} = \frac{\tilde{h}_{k} + \tilde{h}_{l}}{\tilde{h}_{k}\,\tilde{h}_{l}}, \quad
  \beta_2^{k,l} = -\frac{\tilde{h}_{k}}{\tilde{h}_{l}(\tilde{h}_{k} + \tilde{h}_{l})}.
\end{align}
The resulting finite-difference matrix
with respect to the $p$\,-\,th spatial direction is then given by
\begin{align}
 J^p_n :=
 \left( \begin{array}{c c c c c c}
    \beta_{0}^{1,2} & \beta_1^{1,2} & \beta_2^{1,2}  &   &  & \\*[\jot]
    \alpha_{0}^{1,2} & \alpha_{1}^{1,2} & \alpha_{2}^{1,2} &  &  & \\*[\jot]
                     & \alpha_{0}^{2,3} & \alpha_{1}^{2,3} & \alpha_{2}^{2,3} & & \\*[\jot]
      & \quad \quad \,\ddots & \quad \quad \,\ddots & \quad \quad \,\ddots & \quad \quad\,\, & \\*[2\jot]
       &    &  \alpha_{0}^{n-3,n-2} &  \alpha_1^{n-3,n-2} &  \alpha_2^{n-3,n-2} & \\*[\jot]
       &    &    &  \alpha_{0}^{n-2,n-1} &  \alpha_1^{n-2,n-1} &  \alpha_2^{n-2,n-1} \\*[\jot]
       &    &    & -\beta_2^{n-1,n-2} & -\beta_1^{n-1,n-2} & -\beta_{0}^{n-1,n-2} 
 \end{array}\right) \in \mathbb{R}^{n \times n}.
\end{align}
%

\Paragraph
The tensor $\Phi$ representing the scalar potential on the center points of the field cells,
is given by the entries $\phi(x_{\textbf{i}}^{c}) \ \hat{=} \ \phi_{i_1\,i_2\,i_3}$,
and the stray field can now be computed by evaluating the $3$-component tensor
\begin{align} \label{Hfield}
  \textbf{H}_d = -\!\left(
   \begin{array}{c}
    \Phi \times_1 J^1_n \\*[\jot]
    \Phi \times_2 J^2_n \\*[\jot]
    \Phi \times_3 J^3_n 
   \end{array} \right).
\end{align}
Furthermore, the demagnetizing energy is given by~\,\cite{kronmuller_handbook_2007}
\begin{align}
  E_{\text{demag}} = -\frac{\mu_{0}}{2} \int_{\Omega} \textbf{H}_{d} \cdot \textbf{M} \ d^{\,3} x.
\end{align}
On a uniform mesh, a rough estimate for $ E_{\text{demag}} $ is obtained by midpoint quadrature,
\begin{align} \label{energy_approx}
  E_{\text{demag}} \approx \frac{\mu_{0}}{2}\,V_{\text{cell}}
  \sum_{p=1}^{3} \sum_{\textbf{j}}  \textbf{M}^{(p)} \ast (\Phi \times_{p} J^{p}_{n}),
\end{align}
where $V_{\text{cell}}=1/n^3$, and $\ast$ denotes the \textit{Hadamard tensor product}~\cite{kolda_tensor_2009}.
In the case of a non-uniform mesh, the midpoint approximation reads
\begin{align} \label{energy_approx2}
  E_{\text{demag}} \approx \frac{\mu_{0}}{2}
  \sum_{p=1}^{3} \sum_{\textbf{j}} \textbf{V} \ast \textbf{M}^{(p)} \ast (\Phi \times_{p} J^{p}_{n}),
\end{align}
where $\textbf{V}$ denotes the \textit{volume tensor}\, containing the volumes of the computational cells as entries.

\Paragraph
In the case where the $p$-component magnetization tensors are given in canonical tensor format, the functions $G^{(p)}$ from Eq.~\eqref{tuck2}
are in the same format and therefore, from the additive structure of Eqs.~\eqref{quad} and~\eqref{scpot3},
one recognizes $\Phi$ to be given in canonical format (see Eq.~\eqref{cppot} and \ref{tensors3}).
In this case efficient mode multiplication can be applied in~\eqref{Hfield},
see \cite{bader_efficient_2008} for details.
The components of the stray field are then in canonical format,
which enables evaluation of~\eqref{energy_approx} by using inner products , i.e.
\begin{align}
   \sum_{\textbf{j}} \textbf{M}^{(p)} \ast \textbf{H}^{(p)}_d =
   \left\langle \text{vec}(\textbf{M}^{(p)}), \text{vec}(\textbf{H}^{(p)}_d ) \right\rangle,
\end{align}
where $\text{vec}(\cdot)$ denotes vectorization.
The inner product $\left\langle \cdot,\cdot \right\rangle$ can be performed in canonical format at a cost of merely
$O(R_M R_H\,n)$ operations; see \ref{tensors3} and \cite{bader_efficient_2008} for details.
(Here, $R_M,R_H$ denote the Kronecker ranks of $\textbf{M}^{(p)}$ and $\textbf{H}^{(p)}_d$.) 
Additionally Eq.~\eqref{energy_approx2} can be carried out in CP-format since the volume tensor for tensor product grids can be considered as a rank-$1$ tensor. 

\section{Numerical results}
\label{Num}

\Startsec
We use MATLAB~v\,7.11 for our computations, including the Tensor Toolbox~\cite{bader_algorithm_2006}.
All timings are reported for a Linux Workstation with a Quad-Core Intel i7 processor and 6 GB RAM.
\subsection{Accuracy of the magnetic scalar potential}\label{accuracy}
\Startsec
Our approach for computing the scalar potential yields formulae \eqref{sumpot} and \eqref{cppot}, respectively. As indicated in the previous section this has \textit{almost linear} effort, 
i.e. $O(N^{4/3})$, in the general case and \textit{sublinear} effort, i.e. $O(N^{2/3})$, for magnetization in CP format. 
It is notable that our approach leads to the same accuracy as direct integration of Eq.~\eqref{scpot} with complexity $O(N^2)$, i.e.
\begin{align}\label{direct_int}
\phi(\vec{x}) = \frac{1}{4\pi} \sum_{p=1}^{3} \sum_{\textbf{j}} m_{\textbf{j}}^{(p)} \int_{\Omega_{\textbf{j}}} \frac{x^{(p)} - {x^{\prime}}^{\,(p)}}
                              {\left|\vec{x} - \vec{x}^{\,\prime} \right|^3}\,d^{\,3} x^{\prime}.
\end{align}
This is due to the robust \textit{sinc-quadrature} with respect to $\rho$ (see Tab.~\ref{table_sinc}) that we used for representing the kernel function by Eq. \eqref{transform}. \\
\begin{table}
\caption{Comparison of accuracy of Eqs. \eqref{direct_int} and \eqref{sumpot} ($R=50$). Absolute ($Err_{10/50}$) and relative ($Relerr_{10/50}$) $L_2$-errors for a $N=10^3$ (exact errors) and $N=50^3$ (errors for $50$ randomly chosen mesh-points) tensor product grid,
$g^{\text{dir}}$ and $g^{\text{ten}}$ indicate the order of Gaussian quadrature used for the integrals in \eqref{direct_int} and \eqref{Dmat}.} \label{table_L2}
\begin{center}
\begin{tabular}{|c|c||c|c||c|c|}\hline
$g^{\text{dir}}$  & $g^{\text{ten}}$ & $Err_{10}$ & $Relerr_{10}$ & $Err_{50}$  &  $Relerr_{50}$\\ \hline\hline
$16$ & $\,4$ & 5.62\,$e-$05 & 1.35\,$e-$04 & 4.43\,$e-$06 & 4.15\,$e-$04\vph \\ \hline
$16$ & $\,8$ & 2.72\,$e-$08 & 6.55\,$e-$08 & 2.15\,$e-$09 & 2.02\,$e-$07\vph \\ \hline

$\,4$ & $16$ & 5.62\,$e-$05 & 1.35\,$e-$04 & 4.43\,$e-$06 & 4.16\,$e-$04\vph \\ \hline
$\,8$ & $16$ & 2.72\,$e-$08 & 6.55\,$e-$08 & 2.15\,$e-$09 & 2.01\,$e-$07\vph \\ \hline
$16$ & $16$ & 9.33\,$e-$16 & 2.24\,$e-$15 & 2.01\,$e-$16 & 1.89\,$e-$14\vph \\ \hline
\end{tabular}
\end{center}
\end{table}
\begin{table}
\caption{Comparison of accuracy of Eqs. \eqref{direct_int} and \eqref{sumpot} ($R=50$). Absolute ($Err_{10/50}$) and relative ($Relerr_{10/50}$) $L_2$-errors for a $N=10^3$ (exact errors) and $N=50^3$ (errors for $200$ randomly chosen mesh-points) tensor product grid, 
$g^{\text{dir}}$ indicates the order of Gaussian quadrature used for the integrals in \eqref{Dmat}, the
integrals in \eqref{direct_int} are evaluated exactly.} \label{table_L2_2}
\begin{center}
\begin{tabular}{|c||c|c||c|c|}\hline
$g^{\text{ten}}$  & $Err_{10}$  &  $Relerr_{10}$ & $Err_{50}$  &  $Relerr_{50}$\\ \hline\hline
$\,4$ & 2.31\,$e-$04 & 4.52\,$e-$04 & 1.59\,$e-$05 & 4.45\,$e-$04 \vph \\ \hline
$\,8$ & 1.13\,$e-$07 & 2.21\,$e-$07 & 7.77\,$e-$09 & 2.17\,$e-$07 \vph \\ \hline
$16$  & 4.34\,$e-$14 & 8.55\,$e-$14 & 5.85\,$e-$14 & 1.64\,$e-$12 \vph \\ \hline
$32$ & 1.29\,$e-$14 & 2.52\,$e-$14 &  5.86\,$e-$14 & 1.64\,$e-$12 \vph \\ \hline
\end{tabular}
\end{center}
\end{table}

For demonstration we assumed random magnetization uniformly distributed in the interval $(-1,1)$ on a tensor product grid. 
We use the absolute and relative $L_2$-errors as measurements for the accuracy, i.e. 
\begin{align}
	Err_n = & \sqrt{\sum_{\textbf{j}} (\phi_{\textbf{j}}^{\text{dir}} - \phi_{\textbf{j}}^{\text{ten}})^2} \equiv \left\|\Phi^{\text{dir}} - \Phi^{\text{ten}}\right\|_{\text{F}}, \\
	Relerr_n = & \left\|\Phi^{\text{dir}} - \Phi^{\text{ten}}\right\|_{\text{F}} / \left\|\Phi^{\text{dir}}\right\|_{\text{F}},
\end{align}
where $\Phi^{\text{dir}}$ and $\Phi^{\text{ten}}$ are the tensors representing the potential at the center points of the computational cells obtained by the direct
formula \eqref{direct_int} and the tensor approach \eqref{sumpot}, respectively. 
Tab.~\ref{table_L2} shows that both have the same accuracy level for equal orders of Gaussian quadrature for tensor product grids with $N=10^3$ and $N=50^3$. 
Note that the integrands in \eqref{direct_int} get singular for $\vec{x} = \vec{x}^{\prime}$, where $\vec{x}$ denote the centers of field cells. The contribution of the singular integrals 
is, however, zero, since the integrand is an odd function with respect to the $p^{\prime}$-coordinate and so it is possible to treat the singularities with Gaussian quadrature (which is symmetric).   
In Tab.~\ref{table_L2_2} we give the relative and absolute errors for a $N=10^3$ and 
$N=50^3$ grid, where we now evaluate the integrals in \eqref{direct_int} exactly by using the formulas from \cite{kronmuller_handbook_2007}, \textit{Numerical Micromagnetics: Finite Difference Methods}.\\
Essentially there is not any difference in accuracy between the direct integration according to \eqref{direct_int} and the approach in this paper, 
at least for grid-sizes where the used \textit{sinc-quadrature} is guaranteed to be robust with respect to $\rho$, see also Sec.~\ref{tauint}. \\
The same accuracy statement holds for the $O(N^{2/3})$-algorithm since no further approximation is necessary to derive \eqref{cppot} out of \eqref{sumpot}, unless magnetization is given in $\mathcal{C}_{\textbf{n},r}$.  




\subsection{Sublinearity and data-compression in the CP-case}
\Startsec
In order to demonstrate the reduction of computational complexity to sublinearity in the volume size we first assume a constant magnetization distribution in the $z$\,-\,direction of the unit-cube,
i.e., $\textbf{M} \equiv (0, 0, M^{z})$, where $M^{z} = M_{s}\,m_{z} $.
We set the saturation magnetization to $1$ and compare the results with the exact solution,
then given by an energy density of $E_{\text{demag}}/\mu_0 = m_z^2/6$.
\begin{figure}[hbtp]
\centering
\includegraphics[scale=0.5]{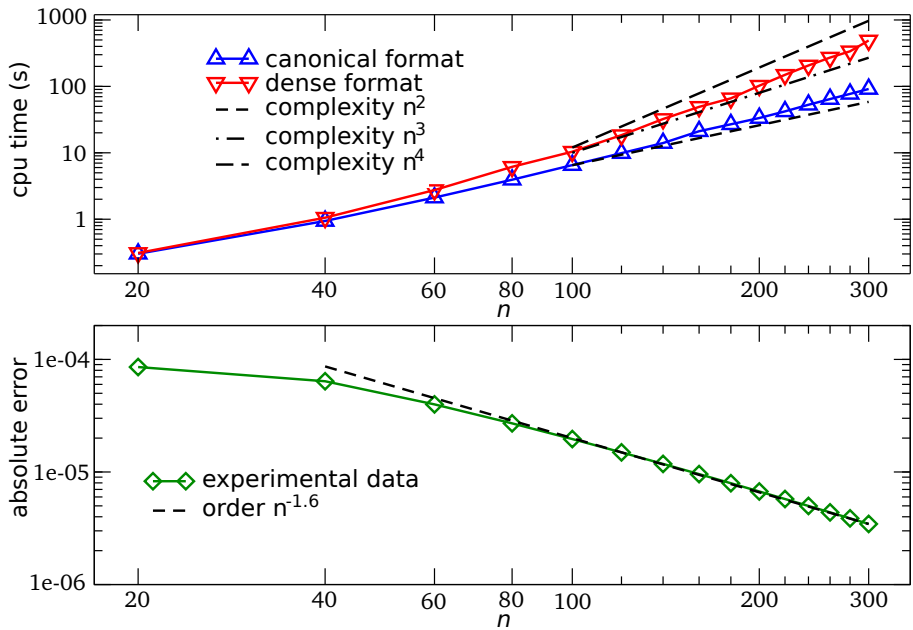}
\caption{
Results for $\textbf{M} \equiv (0, 0, M^{z})$ on the unit cube.
  Absolute error in the energy for $n=20 \hdots 300$ (number of computational cells is $N = n^3$) versus $n$,
  the number of discretization points in one direction.
  In addition, cpu times versus $n$ are plotted for the computation of the scalar potential.
  The curve marked by $ \triangledown $ shows the case with magnetization tensor given in dense tensor format,
  the curve marked by $ \vartriangle $ corresponds to the case with magnetization in canonical format.
  \label{fig1}}
\end{figure}
\Paragraph
Fig.~\ref{fig1} shows the absolute errors in the energy and the cpu times for computation of the scalar potential.
The error decreases with order 
about $1.6$,
the cpu times increase with order $\approx 3.6$ up to $4$
in the case where a dense tensor format is used for the magnetization tensor,
and with an order between $2.4$--$2.6$ (\textit{sublinear}),
when the magnetization is
represented
in canonical tensor format ($r_p = 1, \, p = 1 \hdots 3$). We have used $50$ Gauss-Legendre quadrature terms
and increased the number of sinc terms~$R$ according to $35 + 3n/20$, so using $38$ up to $80$ terms.
\begin{table}
\caption{Computational complexity for calculation of the scalar potential in CP format (randomly assembeled magnetization, i.e. $\textbf{M}^{(p)} \in \mathcal{C}_{n,5}$). Cpu-times (in sec) 
averaged for $20$ identical experiments each and given for increasing $n$ and $R$. In all computations the number of Gaussian-quadrature nodes is $30$.} \label{order}
\begin{center}
\begin{tabular}{|c||c|c|c|c|}\hline
$n$  &  $t_{R=20}$ & $t_{R=40}$ & $t_{R=80}$ & $t_{R=160}$\\ \hline\hline
\, $20$ &  \, 0.24 & \, 0.47 & \, 0.92 & \, 1.84 \vph \\ \hline
\, $40$ &  \, 1.09 & \, 2.10 & \, 4.09 & \, 6.20 \vph \\ \hline
\, $80$ &  \, 3.16 & \, 6.33 & 12.79 & 26.46 \vph \\ \hline
$160$  &  13.46 & 25.68 & 49.85 & 98.74 \vph \\ \hline
$320$  &  50.62 & 97.78 & 189.78 \, & 379.07 \, \vph \\ \hline
\end{tabular}
\end{center}
\end{table}

\Paragraph
For the purpose of verification of the asymptotic operation count of Sec.~\ref{comp}, 
we compute the magnetic scalar potential for randomly assembled $p$-component magnetization tensors of rank $5$, i.e. $\textbf{M}^{(p)} \in \mathcal{C}_{n,5}$, and measure the cpu-times (averaged for $20$ experiments each) with respect to 
increasing $R$ and mesh-parameter $n$. In Tab.~\ref{order} we can observe the linear increase in $N^{2/3} = n^2$ and $R$.    
%

\Paragraph
We next consider a flower-like magnetization state that allows a sufficiently good low rank approximation 
in the CP-format
without severe loss of accuracy. 
The main magnetization direction is taken to be along the $z$\,-\,axis, and the flower is obtained through an in-plane perturbation along the $y$\,-\,axis and an
out-of-plane perturbation along the $x$\,-\,axis.  Assuming polynomial expressions for the perturbations, as in \cite{cowburn_1998}, our flower is the normalized version of
\begin{align} \label{config3}
 M^{x}(r) = & ~ \tfrac{1}{a}(x-x_m)(z-z_m), \nonumber \\
 M^{y}(r) = & ~ \tfrac{1}{c}(y-y_m)(z-z_m) + \tfrac{1}{b^3}\,(y-y_m)^3 (z-z_m)^3,  \\
 M^{z}(r) = & ~ 1 \nonumber,
\end{align}
where $x_m, y_m$ and $z_m$ are the coordinates of the center of the cube. We choose $a=c=0.5$ and $b=1$ and generate a dense magnetization state on a $N=100^3$ tensor product grid with self-energy $1.418772\,e\!-\!01 \,[\mu_{0}^{-1} M_{s}^{-2}]$, 
computed with the optimized parameter $c_0$ for the \textit{sinc-quadrature} (see Sec.~\ref{tauint}) and $50$ nodes for the Gaussian quadrature. \\
We now approximate the above dense magnetization in CP-format by using an alternating least squares algorithm that scales almost linear in $N$, see e.g. \cite{kolda_tensor_2009}. 
Choosing \textit{Kronecker ranks} $r=5$ for each magnetization component, i.e. $\textbf{M}^{(p)} \in \mathcal{C}_{100,5}, \, p=1 \hdots 3$, results in a relative $L_2$-approximation error of less than $1\,e\!-\!06$. 
The data storage requirements for the magnetization tensors have been compressed by a factor of $1.5\,e\!-\!03$. 
For the field computation we use the less accurate \textit{sinc-quadrature} with $R=35$ and only $10$ Gaussian quadrature nodes. Compared to the dense and more accurate computation, 
the CP-approximation only results in a relative error in the energy of $1.8\,e\!-\!04$ and $2.6\,e\!-\!05$ in the relative $L_2$-error norm for the magnetic scalar potential. 
The storage requirements for the stray field are about $13.5\,\%$ of that for the dense case.\\
Computation in the CP-format based on algebraic compression, like in the above example, might not work for arbitrary magnetization and also needs a setup-phase that scales with $O(N^{4/3})$, unless tensor ACA is used like in \cite{goncharov_kronecker_2010}.
CP-based schemes are therefore particulary efficient if the input-magnetization has CP-structure and can be preserved by the algorithm. 
Subsequent work will concentrate on fast and data-sparse static micromagnetic simulations in the space $\mathcal{C}_{\textbf{n},r}$ (minimization in the CP-format), using the tensor approach in this paper for magnetostatic energy computation.

\subsection{Comparison with FEM/BEM}

\begin{table}
\caption{Absolute error in the energy between results for direct tensor integration algorithm and FEM/BEM for uniform magnetization distribution in the unit cube. $N$ indicates the number of total nodes in the mesh. In the fourth column we give
the rel. deviation of the energy values computed by the two numerical schemes; percentage is based on the true value, i.e. $e_d = 1/6\,[\mu_{0}^{-1} M_{s}^{-2}]$.} \label{femme_accuracy}
\begin{center}
\begin{tabular}{|c||c|c|c|c|}\hline
$N$  &  error (tensor) & error (FEM/BEM) & deviation [\%]\\\hline\hline
$15^3$ &  1.38\,$e-$04 & 1.55\,$e-$03 & 8.50\,$e-$1\vph \\ \hline
$30^3$ &  8.19\,$e-$05 & 4.51\,$e-$04 & 3.20\,$e-$1 \vph \\ \hline
$60^3$ &  3.98\,$e-$05 & 3.47\,$e-$04 & 2.32\,$e-$1 \vph \\ \hline
\end{tabular}
\end{center}
\end{table}

Like in the previous section we first compare the proposed scheme of this paper with that obtained by the finite element simulation package FEMME~\,\cite{femme} in the case of uniform magnetization where no dicretization
error for the magnetization arises. The used FEM/BEM implemention solves the weak formulation of the magnetostatic Poisson equation. Dense boundary element matrices are approximated in the \textit{H-matrix format} and preconditioned 
iterative linear solvers are used to gain an almost linear complexity in the volume size, see e.g. \cite{kronmuller_handbook_2007}, \textit{Numerical Methods in Micromagnetics (Finite Element Method)}.\\
From Tab.~\ref{femme_accuracy} one can see that the FEM/BEM algorithm approximates in the case of uniform magnetization configuration about one order of magnitude worse than the direct tensor integration algorithm.
\begin{table}
\caption{Absolute and relative deviation in the energy between results for direct tensor integration algorithm and finite element reference-value for magnetization distribution of a vortex in the unit cube given by Eqs.~\eqref{config1}.
In order to resolve the vortex, we use a non-uniform grid (geometrically refined towards the center of the cube). The columns to the right show the minimal grid size, i.e.\ $\min h_j$,
in the center of the cube and respectively the maximal value, i.e.\ $\max h_j$, next to the boundaries.} \label{table1}
\begin{center}
\begin{tabular}{|c||c|c|c|c|}\hline
$n$  &  abs.\ deviation & rel.\ deviation [\%] & grid-min  & grid-max  \\ \hline\hline
$10$ &  2.02\,$e-$04 & 9.32\,$e-$01 & 8.9\,$e-$02 & 1.1\,$e-$01 \vph \\ \hline
$20$ &  1.61\,$e-$04 & 7.42\,$e-$01 & 3.3\,$e-$02 & 7.2\,$e-$02 \vph \\ \hline
$30$ &  5.58\,$e-$05 & 2.58\,$e-$01 & 1.3\,$e-$02 & 6.6\,$e-$02 \vph \\ \hline
$40$ &  6.65\,$e-$06 & 3.07\,$e-$02 & 5.5\,$e-$03 & 6.6\,$e-$02 \vph \\ \hline
$50$ &  1.78\,$e-$06 & 8.23\,$e-$03 & 2.8\,$e-$03 & 6.4\,$e-$02 \vph \\ \hline
\end{tabular}
\end{center}
\end{table}

\Paragraph
We now take a vortex-like state in a $200\,\text{nm}^3$\,-\,cube, described by the model in \cite{Feldtkeller1965}, i.e.
\begin{align} \label{config1}
 M^{x}(r) = & ~-\frac{y}{r}\,\big( 1 - \exp\big( -4\,\frac{r^2}{r^2_c} \big) \big)^{\frac{1}{2}}, \nonumber \\
 M^{y}(r) = & ~~\frac{x}{r}\,\big( 1 - \exp\big( -4\,\frac{r^2}{r^2_c} \big) \big)^{\frac{1}{2}},  \\
 M^{z}(r) = & ~\exp\big( -2\,\frac{r^2}{r^2_c} \big) \nonumber,
\end{align}
where $r = \sqrt{x^2 + y^2}$, and we choose the radius of the vortex core as $r_c = 28$\,nm.
The vortex center coincides with the center of the cube, and the magnetization is assumed to be rotationally symmetric about the $x=y=100$\,nm axis
and translationally invariant along the $z$\,-\,axis.
\Paragraph
For the above configuration \eqref{config1} and an amount of $50^3$ nodes
and about $5\cdot50^3$ tetrahedral elements, FEMME finds $E_{demag}/\mu_0 = 2.16185\,e\!-\!02$,
which we take as reference value. We compare this value with computations using the direct tensor integration algorithm
introduced in Sec.~\ref{method}
on an adaptive mesh refined geometrically, in each spatial direction, towards the vortex center of the cube,
see Tab.~\ref{table1}.

\Paragraph

\begin{table}
\caption{Absolute and relative deviation between results for direct tensor integration algorithm and finite element reference-value for magnetization distribution in the unit cube given by Eqs.~\eqref{config3}.} \label{table2}
\begin{center}
\begin{tabular}{|c||c|c|c|c|}\hline
$n$  &  abs.\ deviation & rel.\ deviation [\%] \\\hline\hline
$20$ &  3.42\,$e-$04 & 2.24\,$e-$01 \vph \\ \hline
$30$ &  2.83\,$e-$04 & 1.85\,$e-$01 \vph \\ \hline
$40$ &  2.43\,$e-$04 & 1.59\,$e-$01 \vph \\ \hline
$50$ &  2.18\,$e-$04 & 1.43\,$e-$01 \vph \\ \hline
$80$ &  1.83\,$e-$04 & 1.20\,$e-$01 \vph \\ \hline
\end{tabular}
\end{center}
\end{table}

Finally we compare our algorithm with FEMME for a flower-like state of the cube in the previous example, where we choose $a=c=1$ and $b=2$ in Eq.\eqref{config3}.

Using the same finite element mesh as in the previous example,
FEMME now finds $E_{demag}/\mu_0 = 1.52653\,e\!-\!01$,
which we again take 
in order to compare both approximations.
Tab.~\ref{table2} shows absolute and relative deviations,
in this case on a uniform grid used for the direct tensor integration algorithm. 
One can observe a similar difference like in Tab.~\ref{femme_accuracy}.

\section{Conclusions}
\label{conclusions}

\Startsec
We have shown, both theoretically and via numerical experiments, that the algorithm introduced in this paper allows computation of the magnetostatic field and energy in reduced complexity (below linear effort in the number of computational cells used) when magnetization tensors are given in
\textit{canonical format}.
We expect that, in the future, the tensor approximation approach can
be used for computing equilibrium states for a well-defined initial magnetization given in canonical format.

\section*{Acknowledgements}

\Startsec
The authors are grateful to SuessCo KG \cite{femme} for providing the code FEMME for micromagnetic simulations,
which we have used for comparison. The authors also gratefully acknowledge
financial support by the Austrian Science Fund (FWF, project SFB-ViCoM F41).


\appendix

\section{Background on tensor formats}
\label{tensors3}

\Startsec
Here we briefly review some basic facts about tensor formats,
see e.g.~\cite{kolda_tensor_2009}, \cite{bader_efficient_2008} for more details.
Specifically, we consider the case of a $3$-d tensor
$ \textbf{A} \in \mathbb{R}^{n_1 \times n_2 \times n_3} $.

\Paragraph
For a matrix $U \in \mathbb{R}^{m \times n_j}$
the \textit{$j$-mode matrix product} $ \textbf{A} \times_j U $ of the tensor $\textbf{A}$ with $ U $ is
defined element-wise in the following way. E.g.\ for $ j=1 $,
\begin{align} \label{modemul}
(\textbf{A} \times_1 U)_{i_1\,i_2\,i_3} := \sum_{i^{\prime}=1}^{n_1} a_{i^{\prime}\,i_2\,i_3}\,u_{i_1\,i^{\prime}},
\end{align}
i.e., the resulting tensor  $\textbf{A} \times_1 U \in \mathbb{R}^{m \times n_2 \times n_3} $ is obtained
by right-multiplication of the $1$-mode fibers (columns) of $ \textbf{A} $ by $ U $.
Analogously for $ j=2,3 $; the cost for the computation of $ A \times_j U $ is
$O(m \prod_{j=1}^{3} n_j)$ operations in general.

\Paragraph
The tensor $ \textbf{A} $ is said to be in \textit{Tucker format (Tucker tensor)} if it is represented in the form
\begin{align} \label{tucker}
\textbf{A} = \textbf{C} \times_1 U_1 \times_2 U_2 \times_3 U_3,
\end{align}
with the so-called \textit{core tensor} $\textbf{C} \in \mathbb{R}^{m_1 \times m_2 \times m_3}$
and matrices $U_j \in \mathbb{R}^{n_j \times m_j}$.

\Paragraph
The tensor \textbf{A} is said to be in
\textit{canonical format (CANDECOMP/PARAFAC (CP) decomposition)} with \textit{(Kronecker) rank}\, $ R $, if
\begin{align} \label{CP}
\textbf{A} = \sum_{r=1}^{R} \lambda_r\,\,u^{(1)}_r \circ u^{(2)}_r \circ u^{(3)}_r
\end{align}
with $\lambda_r \in \mathbb{R} $,\, unit vectors $u^{(j)}_r \in \mathbb{R}^{n_j}$,
and $\circ$ is the vector outer product.
Abbreviating notation as in \cite{kolda_tensor_2009}, a tensor in CP format is written as
\begin{align} \label{CP2}
\textbf{A} = \llbracket \lambda;\,U^{(1)},U^{(2)},U^{(3)} \rrbracket,
\end{align}
with weight vector $\lambda = [ \lambda_1,\ldots,\lambda_R ] \in \mathbb{R}^{R}$
and matrices $U^{(j)} = \big[\,u^{(j)}_1 \,|\, \hdots \,|\, u^{(j)}_R\,\big] \in \mathbb{R}^{n_j \times R}$.
The storage requirement for the canonical tensor format amounts to $O(R\,\sum_{j=1}^{3} n_j )$.\\
In the following we write $\mathcal{C}_{\textbf{n},r}$ for the set of canonical tensors with mode length $\textbf{n} = (n_1, n_2, n_3)$ and rank $r$, and simple $\mathcal{C}_{n,r}$, when the mode-lengths are equal.

\Paragraph
If the core tensor in Eq.~\eqref{tucker} is given in canonical tensor format,
i.e.\ $\textbf{C} = \llbracket \lambda;\,V^{(1)},V^{(2)},V^{(3)} \rrbracket$ with $V^{(j)}\in \mathbb{R}^{m_j \times R} $,
one can easily transform Eq.~\eqref{tucker} into CP format for a cost of $O(R\,\sum_{j=1}^{3} m_j\,n_j )$ operations, giving
\begin{align} \label{cpviatuck}
\textbf{A} = \llbracket \lambda;\,U^{(1)}V^{(1)},U^{(2)}V^{(2)},U^{(3)}V^{(3)} \rrbracket.
\end{align}

The \textit{inner product} for two canonical tensors $\textbf{A} \in \mathcal{C}_{\textbf{n},r_1}$ and $\textbf{B} \in \mathcal{C}_{\textbf{n},r_2}$ , as well as many other operations, can be performed with reduced complexity 
(for the inner product operation it acounts to $O(r_1 r_2 \sum_j n_j )$ ), see e.g. \cite{bader_efficient_2008}. The reduced complexity (and also the data-sparsity) makes it worth developing algorithms in CP format.

\bibliographystyle{plain}
\bibliography{bibrefs}

\end{document}